\documentclass[12pt]{iopart}

\usepackage{iopams}
\usepackage[dvips]{graphicx}
\begin{document}

\title{Diffusion-annihilation processes in weighted scale-free networks with identical degree sequence}

\author{Yichao Zhang}\address{Department of Computer Science and Technology, Tongji
University, 4800 Cao'an Road, 201804, Shanghai, China}
\author{Zhongzhi Zhang}\ead{zhangzz@fudan.edu.cn}\address{School of Computer Science, Fudan University, 200433,
Shanghai, China\\Shanghai Key Lab of Intelligent Information
Processing, Fudan University, 200433, Shanghai, China}
\author{Jihong Guan}\ead{jhguan@tongji.edu.cn}\address{Department of Computer Science and Technology, Tongji University,
4800 Cao'an Road, 201804, Shanghai, China}
\author{Shuigeng Zhou}\ead{sgzhou@fudan.edu.cn}\address{School of Computer Science, Fudan University, 200433,
Shanghai, China\\Shanghai Key Lab of Intelligent Information
Processing, Fudan University, 200433, Shanghai, China}

\begin{abstract}
The studies based on $A+A \rightarrow \emptyset$ and $A+B
\rightarrow \emptyset$ diffusion-annihilation processes have so far
been studied on weighted uncorrelated scale-free networks and
fractal scale-free networks. In the previous reports, it is widely
accepted that the segregation of particles in the processes is
introduced by the fractal structure. In this paper, we study these
processes on a family of weighted scale-free networks with identical
degree sequence. We find that the depletion zone and segregation are
essentially caused by the disassortative mixing, namely, high-degree
nodes tend to connect with low-degree nodes. Their influence on the
processes is governed by the correlation between the weight and
degree. Our finding suggests both the weight and degree distribution
don't suffice to characterize the diffusion-annihilation processes
on weighted scale-free networks.

\end{abstract}

\section{Introduction}\label{intro}

Complex networks are a powerful and versatile mathematical tool for
representing and modeling the the structure of complex
systems~\cite{RMP7447,AP511079}. Their wide applications in distinct
areas have made them an extensively focused discipline in the past
decade~\cite{SIAMR45167,PR424175}. Prompted by data mining and the
increased computing power of computers, extensive empirical studies
have unveiled that most real networked systems can be characterized
by a power-law degree distribution $P(k)\sim k^{-\gamma}$, leading
to the rising of research on our basic understanding of the
organization of many real-world systems in nature and
society~\cite{RMP7447,AP511079,SIAMR45167,PR424175}. The
characteristic exponent $\gamma$, usually observed in the range
$\in(2,3]$ in recent empirical
studies~\cite{RMP7447,AP511079,SIAMR45167,PR424175} is very
important since it fundamentally influences some dynamical processes
on the scale-free networks, e.g.,
synchronization~\cite{IEEETCS5392,PRE74056116}, disease
spreading~\cite{PRL89108701}, and so forth. Among these processes,
one aspect that has recently received considerable attention is the
diffusion-annihilation problem, i.e., bimolecular chemical and
physical reactions of the identical particles $A+A \rightarrow
\emptyset$ and different particles $A+B
\rightarrow\emptyset$~\cite{PRL92138301,PRE71056104,NJP11063025,PRE82021108,JPA272633,JPA38R79,PRA463132,PRSL387147,JCP782642}.

Unlike diffusion-reaction, the substances in diffusion-annihilation
process don't yield products with mass. In the study of
diffusion-annihilation, the density $\rho$ of the surviving
particles is thus a crucial problem since it presents a quantitative
description of the reaction process. In the large time limit, $\rho$
behaves as
\begin{equation}
\frac{1}{\rho(t)}-\frac{1}{\rho(0)}=k\cdot t^{f},
\end{equation}
where $k$ is the rate constant and $\rho(0)$ is the particle density
at $t=0$. In the mean-field approximation with
$\rho_A(0)=\rho_B(0)$, both processes can be described as
$\frac{d\rho(t)}{dt}=-const\cdot{\rho(t)}^2$, whose solution is
$f=1$. The solution is valid in regular lattices of Euclidean
space~\cite{JPA272633} with a spatial dimension $d>d_c$, where $d_c$
is the critical dimension of this process. For the $A+A \rightarrow
\emptyset$ process, $d_c=2$ while for the $A+B \rightarrow
\emptyset$ process $d_c=4$~\cite{JPA38R79}. Further studies on
fractals found that the exponent $f=\frac{d_s}{4}$ for $A+B
\rightarrow \emptyset$~\cite{PRA463132}, where $d_s$ is the spectral
dimension of the fractal structure.

As the existence of the depletion zone ($A+A \rightarrow
\emptyset$)~\cite{PRSL387147} and segregation of the reactants ($A+B
\rightarrow \emptyset$)~\cite{JCP782642}, the upper bound of the
exponent $f$ for the regular lattices is $1$. Whereas, when the
processes are performed on scale-free networks with identical nodes
and links, $f$ can be considerably higher than
$1$~\cite{PRL92138301}. Inspired by the observations, the relation
between $\gamma$ and $f$ on $A+A \rightarrow \emptyset$ was
investigated analytically~\cite{PRE71056104} in uncorrelated
scale-free networks~\cite{PRL87258701,PRE65066130}. Put briefly, the
term ``uncorrelated" denotes that no degree-degree correlations
among nodes exist in the networks, namely, the conditional
probability $P(k'|k)$ that a node of degree $k$ is connected to a
node of degree $k'$ can be formalized as $\frac{k'P(k')}{\langle
k\rangle}$. The analytical solution shows $f$ is only governed by
the exponent $\gamma$ for this class of scale-free networks.
Subsequently, an interesting study of $A+B \rightarrow \emptyset$ on
fractal scale-free networks shows the segregation can also be found
in the scale-free networks~\cite{NJP11063025}. Influenced by the
segregation, the reaction process is hampered apparently.

Very recently, considering heterogeneous distributions of
weights~\cite{PNAS1013747,PRE80016107}, a heuristic research on the
weighted uncorrelated scale-free networks analytically present a
more realistic conclusion~\cite{PRE82021108}. In this work, the
weight of links is defined as $w_{ij}=(k_ik_j)^{\theta}$ with the
degree $k_i$ and $k_j$ of both nodes, where $\theta$ is the
network's weightiness parameter which characterizes the dependence
between link weight and the node degrees~\cite{PNAS1013747}. When
$\theta = 0$, there is no dependence between link weight and node
degree, all link weights are equal with one, and the network becomes
an unweighted network. When $\theta > 0$, it is a weighted network
where links have different weights. The larger $\theta$ and the
wider difference between links. Based on the mean-field rate
equation for the average density $\rho_k$ of a node with degree $k$,
for the $A+A \rightarrow \emptyset$ process, the authors showed
\begin{eqnarray}
  \rho &\sim& \left\{ \begin{array}{ccc}
      t^{-1}  & \qquad & \theta<\frac{\gamma-3}{2}\\
      t^{-\frac{1+\theta}{\gamma-\theta-2}} & \qquad &
      \frac{\gamma-3}{2}\leq\theta<\gamma-2\\
      e^{-t} & \qquad & \theta\geq \gamma-2
    \end{array} \right., \label{AA}
\end{eqnarray}
in asymptotically large networks. For the $A+B \rightarrow
\emptyset$ process, inserting the mapping relation, they claimed
\begin{eqnarray}
  \rho &\sim& \left\{ \begin{array}{ccc}
      t^{-1}  & \qquad & \theta<\frac{\gamma-3}{2}\\
      (t\ln t)^{-1} & \qquad & \theta=\frac{\gamma-3}{2}\\
      t^{-\frac{1}{\gamma-\theta-2}} & \qquad & \frac{\gamma-3}{2}<\theta<\gamma-2
    \end{array} \right..\label{AB}
\end{eqnarray}
It has been shown that, $f$ is only governed by the weight and
degree distribution.

In this paper, we study a family of weighted scale-free networks
with the identical degree sequence (weighted IDS-SF networks), the
reaction processes are vastly different from the previous
reports~\cite{PRL92138301,PRE71056104,NJP11063025,PRE82021108}. To
this end, we briefly introduce weighted random diffusion in
Section~\ref{sec:1}. Section~\ref{sec:2} is devoted to explicit the
IDS-SF networks. In Section~\ref{sec:3}, our extensive numerical
simulations are compared with previous analytic results of the
diffusion-annihilation processes running on top of the weighted
uncorrelated scale-free networks~\cite{PRE82021108}. Finally, our
conclusions are presented in Section~\ref{sec:conclusion}. Our
findings indicate that the disassortative mixing of the nodes is the
essential reason for generation of the depletion zone and
segregation in this class of scale-free networks.

\section{WEIGHTED RANDOM DIFFUSION}\label{sec:1}

Before introducing the construction of the networks, we briefly
introduce the general random walk on weighted networks to clarify
the influence of high-degree nodes (hubs) on the weighted random
diffusion on scale-free networks. Random walk is a mathematical
formalization of a trajectory that consists of taking successive
random steps. A familiar example is the random walk phenomenon in a
liquid or gas, known as Brownian motion~\cite{AP17549,AP21756}.
Random walk is also a fundamental dynamic process on complex
networks~\cite{PRL92118701}. Random walk in networks has many
practical applications, such as navigation and search of information
on the World Wide Web and routing on the
Internet~\cite{PRE63041108,PRE64046135,PRL89248701,PRE74046118,PA385743}.

Let's consider a weighted random walker starting from node $i$ at
step $t=0$ and denote $P_{im}(t)$ as the probability of finding the
walker at node $m$ at step $t$. The probability of finding the
walker at node $j$ at the next step is
\begin{equation}
P_{ij}(t+1)=\sum_{m}a_{mj}\cdot\Pi_{m\rightarrow j}\cdot
P_{im}(t),\label{pi}
\end{equation} where $a_{mj}$ is an element of
the network's adjacent matrix.

In this case, we define the weight of a link between nodes $i$ and
$j$ as
\begin{equation}\label{wij}
 w_{ij}=w_{ji}=\left\{
 \begin{array}{cl}
0 & \mbox{link i-j doesn't exist}  \\[0.5cm]
(k_ik_j)^\theta & \mbox{link i-j exists}~
\end{array}
\right.,
\end{equation}
where $k_i$ and $k_j$ denote the degree of node $i$ and $j$
respectively. On the other hand, the strength of node $i$ is defined
as
\begin{equation}
s_i=\sum_{j\in\Gamma(i)}w_{ij}=\sum_{j\in\Gamma(i)}{(k_ik_j)^{\theta}},
\label{si}
\end{equation}
Thus the probability $P_{ij}(t)$ for the walker to travel from node
$i$ to node $j$ in $t$ steps is
\begin{eqnarray}
P_{ij}(t)&=&\sum_{m_1,...,m_{t-1}}\frac{w_{im_1}}{s_i}\times
\frac{w_{m_1m_2}}{s_{m_1}}\nonumber\\ &\times&\ldots\times
\frac{w_{m_{t-1}j}}{s_{m_{t-1}}}.
\end{eqnarray}
In other words,
$P_{ij}(t)=\sum_{m_1,...,m_{t-1}}P_{im_1}P_{m_1m_2}\cdot\cdot\cdot
P_{m_{t-1}j}$. Comparing the expressions for $P_{ij}$ and $P_{ji}$
one can see that $s_iP_{ij}(t)=s_jP_{ji}(t)$. This is a direct
consequence of the undirectedness of the network. For the stationary
solution, one obtains $P_i^\infty={s_i}/{Z}$ with $Z=\sum_is_i$.
Note the stationary distribution is, up to normalization, equal to
$s_i$, the strength of the node $i$. This means the higher strength
a node has, the more frequently it tends to be visited by a walker.
Notably, for degree uncorrelated networks~\cite{PRE74026121}, $s_i$
in the steady state scales with $k_i$ as $s_i\sim
k^{\theta+1}_i$~\cite{PRE80016107}.

\section{The scale-free networks with identical degree sequence (IDS-SF networks)}\label{sec:2}
\begin{figure}
\begin{center}
\scalebox{0.8}[0.8]{\includegraphics[width=0.6\linewidth,trim=0 0 0
0]{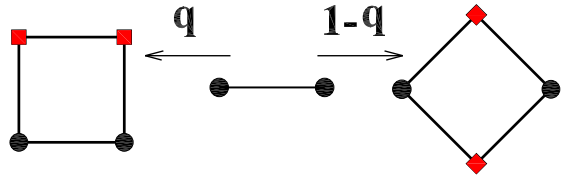}} \caption{(Color online) Iterative method of
the network construction. Each edge is replaced by the connected
clusters on the left-hand side with a certain probability $q$,
otherwise by the one on the right-hand side, where red squares
represent new nodes.}\label{fig1}
\end{center}
\end{figure}

\begin{figure}
\begin{center}
\scalebox{0.8}[0.8]{\includegraphics[trim=0 0 0
0]{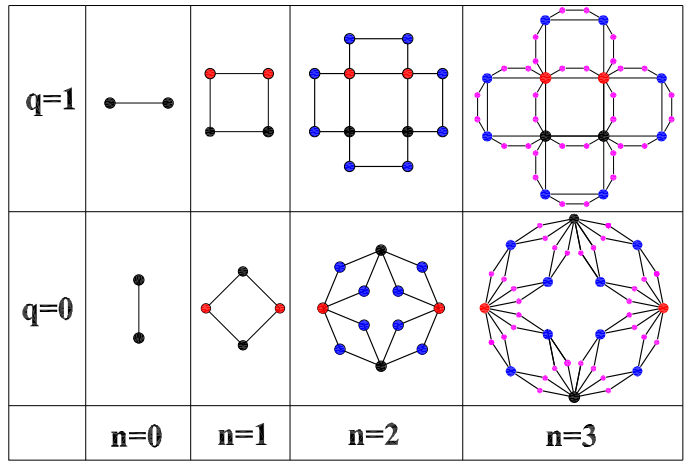}} \caption{(Color online) Illustration
of the first three iterations of the network for the particular
cases $q=0$ and $q=1$.} \label{illustration}
\end{center}
\end{figure}

The scale-free networks with identical degree sequence are a common
topic in complex networks, which offer researchers a platform to
understand how the dynamical behaviors are influenced by the degree
heterogeneity of networks~\cite{AL6161,PRE71027103}. As a class of
the these networks~\cite{PRE79031110,PRE80061111}, the construction
of the present model is controlled by a parameter
$q$~\cite{PRE79031110,PRE80061111} as shown in Fig.~\ref{fig1},
evolving in a recursive way. We denote the network after $n$
iterations by $G(n)$, $n\geq 0$. Then the networks are constructed
as follows. For $n=0$, the initial network $G(0)$ consists of two
nodes connected to each other by a link. For $t \geq 1$, $G(n)$ is
obtained from $G(n-1)$. That is to say, to obtain $G(n)$, one can
add three links to each link existing in $G(n-1)$ (as shown on the
left of Fig.~\ref{fig1}) with probability $q$, or replace it with a
quadrangle (as shown on the right of Fig.~\ref{fig1}) with
complementary probability $1-q$. In Fig.~\ref{illustration}, next,
we present the first three iterations of two special networks
corresponding to two limiting cases $q=0$ and $q=1$, respectively.

\begin{figure}
\begin{center}
\scalebox{0.9}[0.9]{\includegraphics[trim=0 0 0 0]{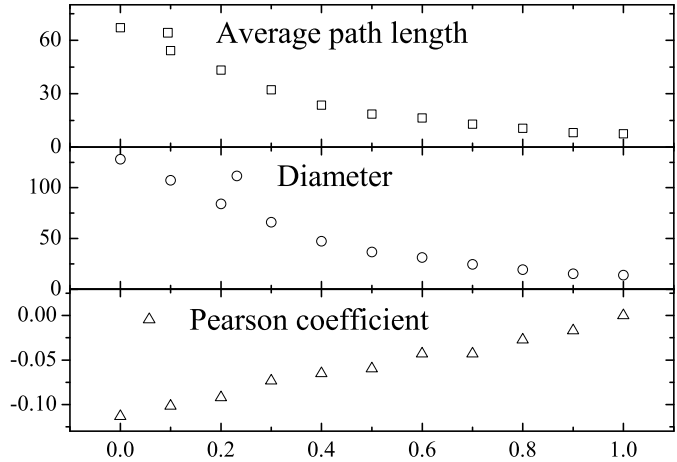}}
\caption{ Pearson correlation coefficient, average path length, and
diameter versus $q$ ranging from $0$ to $1$ for the IDS-SF networks.
Each data point corresponds to ten independent realizations of the
network for $n=7$.}\label{pearson}
\end{center}
\end{figure}

As discussed in the reference~\cite{PRE79031110}, these two limiting
cases and the middle cases ($0<q<1$) exhibit many interesting
properties. For instance, the same degree sequence independent of
parameter $q$, the identical degree distributions, and no
triangles~\cite{SIAM452167} formed by connections among the
neighbors. Note that, as shown in Fig.~\ref{pearson} the Pearson
coefficient increases with $q$ generally, indicating the IDS-SF
networks are disassortative for $q=0$ (the index tends to $-0.5$ as
$N \rightarrow\infty$~\cite{NJP9175}) and uncorrelated for
$q=1$~\cite{PRL89208701}. Hence, for $q=1$, the topological
structure of network satisfies the conditions of applying
$P(k|k')=kP(k)/\langle k\rangle$~\cite{PRE82021108} and mean field
approximation well, which will be discussed in Section~\ref{CASE 1}
in detail. Adopting several $q$ values from $0$ to $1$, one can
generate various networks, for example, fractal ($q=0$) and
non-fractal ($q=1$) networks. These particular features have the
kinetics taking place upon the model be distinct from the well known
results for other networks, e.g., the Barab\'asi-Albert (BA)
graph~\cite{PRE71056104,SCI286509} and uncorrelated configuration
networks~\cite{PRE71056104,PRE71027103}. In the following, we will
show a number of interesting behaviors of Diffusion-annihilation
processes on the networks.
\section{Diffusion-annihilation processes on the weighted IDS-SF networks}\label{sec:3}
According to the conclusion on the weighted random diffusion
operating in the weighted uncorrelated scale-free networks in
Section~\ref{sec:1}, it is easily seen that
$P_i^\infty=\frac{k^{\theta+1}_i}{\Sigma_i k^{\theta+1}_i}$. Thus,
for $\theta>0$, particles move towards hubs with time gradually. As
hubs are the minority of the population, moving to them means
getting concentrated actually. At these hubs, particles have thus a
high probability to collide and react with each other, leading to a
higher reaction rate than that in homogeneous
networks~\cite{PRE71056104}. For $\theta<0$, conversely, the
particles are repelled by the hubs. In this case, the particles are
getting dispersed on the low-degree nodes with time, which are also
called leaves. Seen in this light, reaction rate of
diffusion-annihilation tends to decrease with $\theta$ for the two
processes. In what follows, we will show the diffusion tendency
mentioned above is correct, but the influence of $\theta$ on the
reaction rate is not monotonic, which depends not only on degree
distribution but also other topological features of the network.

We first generate a special IDS-SF structure through an iterative
way with $n=7$. The simulation results are obtained on IDS-SF
networks with $10,924$ nodes and $16,384$ links. For the two
reaction processes, each node in the networks can host at most one
particle. The concrete processes are defined as follows: an
arbitrary particle jumps with a certain probability
$\frac{w_{ij}}{s_i}$ from a node $i$ to a randomly chosen nearest
neighbor $j$. If it is empty, the particle fills it, leaving $i$
empty. If $j$ is occupied, the two particles annihilate, leaving
both nodes empty. An initial fraction $\rho(0)$ of nodes in the
networks is randomly chosen, which is occupied by an $A$ particle
with probability $0.5$ for both types. For the $A+B \rightarrow
\emptyset$ process, the initial densities of $A$ and $B$ are equal,
i.e., $\rho_B(0)=\rho_A(0)$. For a convenience of discussion, we
define $f$ as the first order derivative of $\frac{1}{\rho(t)}$,
where $\rho(t)=\rho_A(t)$ for $A+A \rightarrow \emptyset$ and
$\rho(t)=\rho_A(t)+\rho_B(t)$ for the $A+B \rightarrow \emptyset$
process. In the cases among $0<q<1$, each plot corresponds to $100$
simulations that are ten runs for ten independent realizations of
the network with the same parameters. For the two limiting cases
$q=0,1$, each plot corresponds to $100$ runs for the two
deterministic networks.

As the degree sequences of the IDS-SF networks with $q\in[0,1]$ are
the same, in which $\gamma=3$~\cite{PRE79031110}, Eq.~(\ref{AA}) and
Eq.~(\ref{AB}) can be rewritten as
\begin{eqnarray}
  \rho &\sim& \left\{ \begin{array}{ccc}
      t^{-1}  & \qquad & \theta<0\\
      t^{-\frac{1+\theta}{1-\theta}} & \qquad &
      0\leq\theta<1\\
      e^{-t} & \qquad & \theta\geq 1
    \end{array} \right.. \label{IDS-SF_AA}
\end{eqnarray}
For the $A+B \rightarrow \emptyset$ process, inserting $\gamma=3$,
one can also obtain
\begin{eqnarray}
  \rho &\sim& \left\{ \begin{array}{ccc}
      t^{-1}  & \qquad & \theta<0\\
      (t\ln t)^{-1} & \qquad & \theta=0\\
      t^{-\frac{1}{1-\theta}} & \qquad &
0<\theta<1
    \end{array} \right..\label{IDS-SF_AB}
\end{eqnarray}

For a convenience, we define two quantities as follow:
\begin{equation}
Q_{AA}=\frac{N_{AA}(t)}{M(t)(M(t)-1)},
\end{equation}
\begin{equation}
Q_{AB}=\frac{N_{AB}(t)}{M(t)(M(t)-1)},
\end{equation}
where $N_{AA}(t)$ denotes the number of close contacts between two
nodes with the identical particles for the $A+A \rightarrow
\emptyset$ process. $N_{AB}(t)$ denotes the number of contacts
between the distinct particles for the $A+B \rightarrow \emptyset$
process at time $t$~\cite{JPCM19065123}. $M(t)$ denotes the total
number of particles at time $t$.

\subsection{Case of $q=1$}\label{CASE 1}

As shown in Fig.~\ref{illustration}, in the case $q=1$, the networks
are reduced to the $(1,3)$-flower proposed in the
reference~\cite{NJP9175}. By definition~\cite{DRFDS}, the fractal
dimension $d_f$ can be obtained by
\begin{equation}
d_f=\lim_{n\rightarrow\infty}\left(\frac{\ln{N_n}}{\ln{l_n}}\right),\label{d_f}
\end{equation}
where $N_n$ and $l_n$ are the size and diameter of $G_n$
respectively. Inserting $N_n=\frac{2}{3}(4^n+2)$ and
$l_n=2n$~\cite{NJP9175} into Eq.~(\ref{d_f}), we have
\begin{equation}
d_f=\lim_{n\rightarrow\infty}\left(3\ln2n\right).
\end{equation}
Obviously, the net is infinite-dimensional, namely, a non-fractal
network.
\begin{figure}
\begin{center}
\scalebox{1.4}[1.2]{\includegraphics[trim=0 0 0
0]{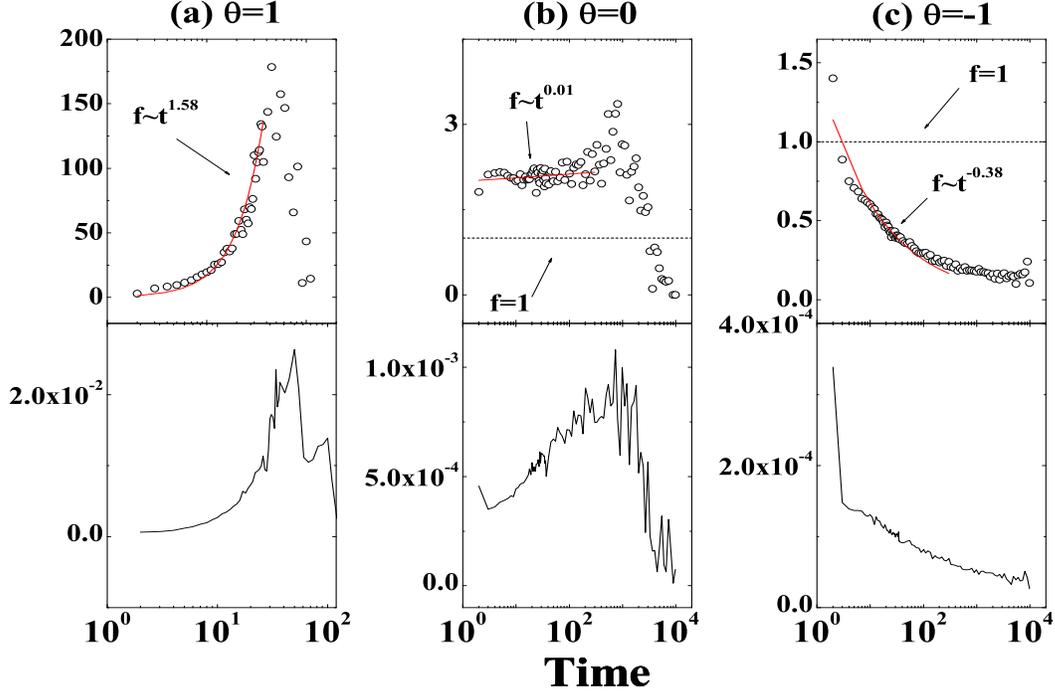}}\caption{(Color online) $f$ and $Q_{AA}$ as a
function of time $t$ for $A+A\to\emptyset$ with $q=1$. The first and
second row denote $f$ and $Q_{AA}$ versus $t$ respectively. The
dashed lines correspond to the mean field prediction: For the
$A+A\to\emptyset$ process, $f=1$ when $\theta=0,-1$, and $f=exp(t)$
when $\theta=1$. }\label{AA_q=1}
\end{center}
\end{figure}
For $A+A \rightarrow \emptyset$, Fig.~\ref{AA_q=1} shows the
relation between $f$ and time $t$ for $\theta=-1,0,1$, where red
lines are the power-law fittings of the plots. Note that all the
plots about the dependence of $f$ and $Q_{AA(AB)}$ on time $t$ are
logarithmically binned in this paper. Concretely,
$f_{t_2}=\frac{\frac{1}{\rho(t_2)}-\frac{1}{\rho(t_1)}}{t_2-t_1}$
and $Q_{t_2}=\frac{Q_{t_2}-Q_{t_1}}{t_2-t_1}$, where the time
interval $log(t_2)-log(t_1)=0.1$. For each panel, the curve is
relatively stable in the beginning and fluctuates radically in the
end. The numerical results show the reaction processes are vastly
different from the previous analytical predictions on the weighted
uncorrelated networks denoted by the dashed
lines~\cite{PRE82021108}.

Compared with the mean-field prediction, one can observe many
discrepancies in Fig.~\ref{AA_q=1}. Here, we only focus on looking
for some common reasons. Note that, for $\theta\geq1$, the
mean-field prediction $f=e^t$ is much higher than our results in
Fig.~\ref{AA_q=1}(a). To show the plots clearly, we omit the dashed
line in this panel. The apparent discrepancy is mainly caused by the
approximation $N_g\rightarrow\infty$. In this condition, $\langle
k^{1+\theta}\rangle\rightarrow\infty$, which makes the differential
equation solvable. For $0<\theta<1$, on the other hand, the
approximation in the literature~\cite{PRE82021108} omits the
reaction running on the low-degree nodes, which causes the predicted
reaction rate is lower than our observation. For $\theta=0$, our
results in Fig.~\ref{AA_q=1}(b) roughly match the conclusion in the
reference~\cite{PRE82021108} in term of the scaling of $f$. But, the
value of $f$ is a bit higher than the prediction in that the global
mean first-passage time of random walks $G$ in the mean field
prediction of Eq.~(\ref{IDS-SF_AA}) is proportional to
$N_g$~\cite{PRE80065104} while $G\sim
{N_g^{\frac{ln3}{ln4}}}$~\cite{PRE80061111}. So that, one can expect
a larger deviation in the IDS-SF networks with $n>7$. Notably, this
observation in this case is inconsistent with the previous
conclusion on finite size effects, i.e., $\frac{1}{\rho(t)}\sim
N^{\frac{3-\gamma}{2}}t$ for $\gamma\leq3$~\cite{PRE71056104}. For
$\theta<0$, our results in Fig.~\ref{AA_q=1}(c) are basically lower
than the prediction. This deviation is caused by the approximation
in Taylor expansion. As is known, $\rho(t)$ can only be omitted at
the end of the reaction, where it is close to $0$.

\begin{figure}
\begin{center}
\scalebox{1.4}[1.2]{\includegraphics[trim=0 0 0
0]{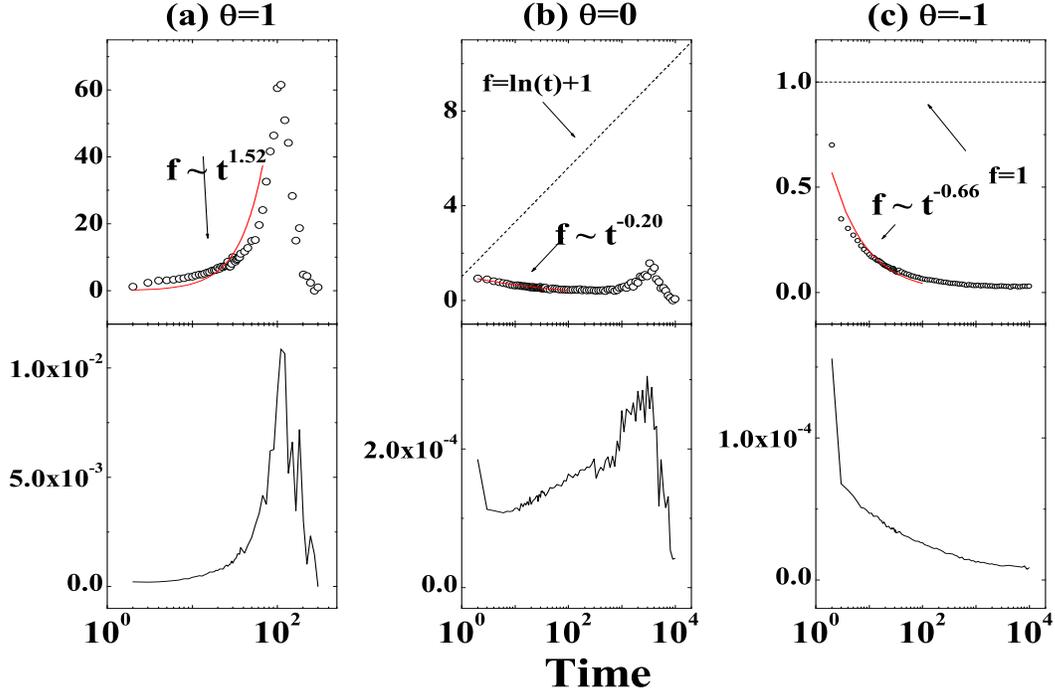}}\caption{(Color online) $f$ and $Q_{AB}$ as a
function of time $t$ for $A+B\to\emptyset$ with $q=1$. The first and
second row denote $f$ and $Q_{AB}$ versus $t$ respectively. The
dashed lines correspond to the mean field prediction: For
$A+B\to\emptyset$, $f=1$ when $\theta=-1$ and $f=ln(t)+1$ when
$\theta=0$. }\label{AB_q=1}
\end{center}
\end{figure}
For the $A+B \rightarrow \emptyset$ process, we measure the relation
between the total particle density $\rho(t)=\rho_A(t)+\rho_B(t)$ and
time $t$ as shown in Fig.~\ref{AB_q=1}. Our observation exhibits the
similar behaviors with $A+A \rightarrow \emptyset$. As the
probability of collision between two identical particles is equal to
that for distinct ones, one can find $Q_{AB}$ in Fig.~\ref{AB_q=1}
is about half of the corresponding $Q_{AA}$. Thus, the reaction rate
of the $A+B \rightarrow \emptyset$ process is naturally much lower
than that of $A+A \rightarrow \emptyset$.

It should be mentioned that the $(1,3)$-flower is a network with a
number of common properties, e.g., non-fractal topology, no degree
correlations and scale-free degree distribution, satisfying the
condition of mean-field approximation fully. But, the annihilation
dynamics on it presents many unpredictable properties. Thus, there
is a need to provide such a complement to the previous discussion on
both fractal scale-free networks~\cite{NJP11063025} and weighted
scale-free networks~\cite{PRE82021108}. Without loss of generality,
in what follows, we will investigate the other limiting case $q=0$.

\subsection{Case of $q=0$}\label{CASE 0}
Unlike the case of $q=1$ addressed above, for $q=0$, the networks
are reduced to the $(2,2)$-flower as shown in the corresponding
panel of Fig.~\ref{illustration}, which is a fractal network whose
fractal dimension $d_f=\frac{\ln4}{\ln2}=2$~\cite{NJP9175}. By
definition, the fractal network is a network satisfying the fractal
scaling $N_B(l_B)\sim l_B^{d_f}$, where $N_B$ is the number of boxes
needed to cover the entire network with boxes of size $l_B$. Note
that the fractal scaling $d_f$ holds in the system where hubs are
located separately from each other~\cite{PRE72045105,NP72045105}. As
is known, the mean-field theory can only be applicable when the nets
have infinite dimensionality but not in the fractal
ones~\cite{NJP9175,NJP11063025}. Thus, the discrepancies between the
mean field prediction and our results are not unexpected. However,
the weighted networks have their unique subtle properties, which
gives rise to many interesting dynamical behaviors distinct from the
previous unweighted fractal nets.

\begin{figure}
\begin{center}
\scalebox{1.4}[1.2]{\includegraphics[trim=0 0 0
0]{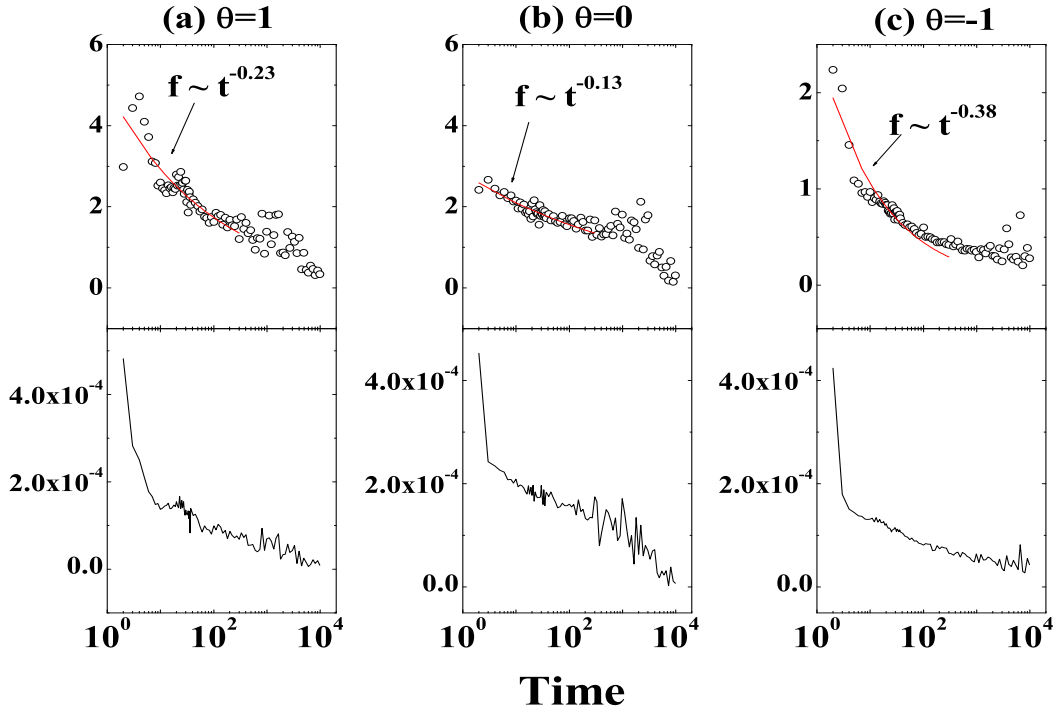}}\caption{(Color online) $f$ and $Q_{AA}$ as a function of
time $t$ for $A+A\to\emptyset$ with $q=0$. The first and second row
denote $f$ and $Q_{AA}$ versus $t$ respectively.}\label{AA_q=0}
\end{center}
\end{figure}

For the $A+A\rightarrow \emptyset$ process in Fig.~\ref{AA_q=0}, we
also measure the relation between $f$ and $t$ for the set of
$\theta$. In Fig.~\ref{AA_q=0}, one can observe that $f$ decays with
time $t$ in all the three panels. In the case of $\theta=1$ in
Fig.~\ref{AA_q=0}(a), the exponent $f$ decreases with $t$ abnormally
and exhibits a contrary behavior with the case of $q=1$, in which
$f$ increases with $t$.
\begin{figure}
\begin{center}
\scalebox{1.4}[1.2]{\includegraphics[trim=0 0 0
0]{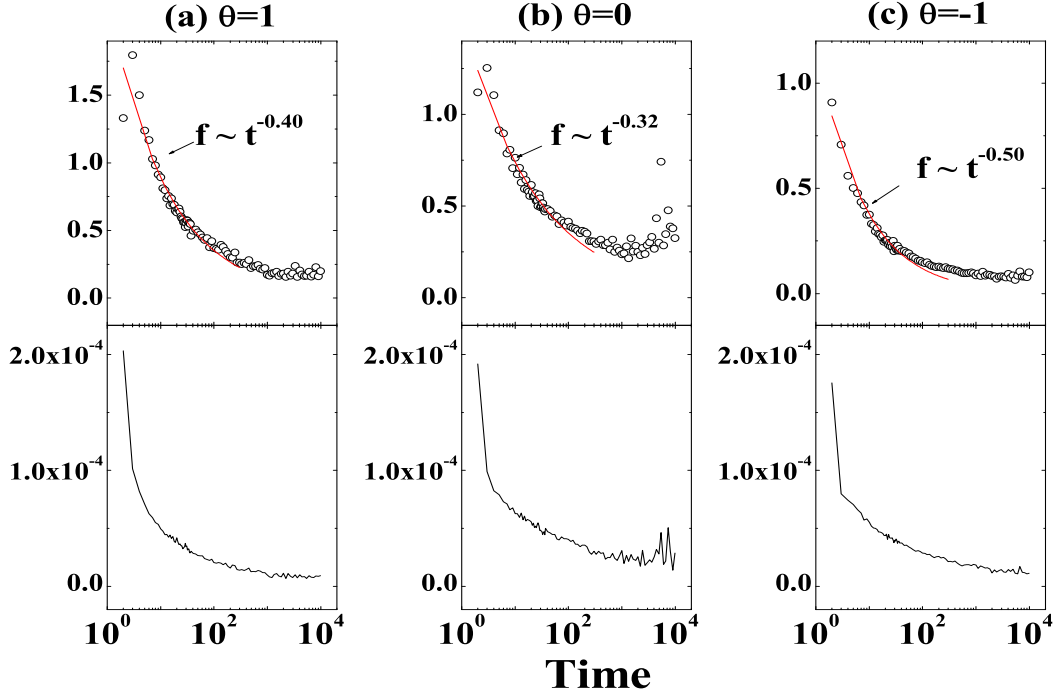}}\caption{(Color online) $f$ and $Q_{AB}$ as a function of
time $t$ for $A+B\to\emptyset$ with $q=0$. The first and second row
denote $f$ and $Q_{AB}$ versus $t$ respectively.}\label{AB_q=0}
\end{center}
\end{figure}
For the $A+B\rightarrow \emptyset$ process shown in
Fig.~\ref{AB_q=0}, one can observe a similar phenomenon with the
$A+A\rightarrow \emptyset$ process as well. Because of
$Q_{AA}\sim2Q_{AB}$, $f$ in this case is much lower than the
$A+A\rightarrow \emptyset$ as well. Notably, for the unweighted
case, i.e., $\theta=0$, $f$ shown in Fig.~\ref{AB_q=0}(b) is not a
constant $0.5$ mentioned in the reference~\cite{NJP11063025}.

For homogeneous initial distributions with equal densities of $A$
and $B$, $\rho_A(0)=\rho_B(0)$, local hubs and the random
fluctuation in the initial particle number generate the segregation
of distinct particles, which drastically slows down the reaction
rate. Usually, for the unweighted uncorrelated scale-free networks,
it is hard for a large number of particles to form a close formation
that cannot be penetrated by the other species because of a short
diameter. However, for $q=0$, the influence of disassortative mixing
is enhanced by the high heterogeneous weight distribution as shown
in Fig.~\ref{AA_q=0}(a) and Fig.~\ref{AB_q=0}(a). The tighter local
hubs attract the particles, the lower the diffusion rate is.

\begin{figure}
\begin{center}
\scalebox{0.4}[0.4]{\includegraphics[trim=0 0 0 0]{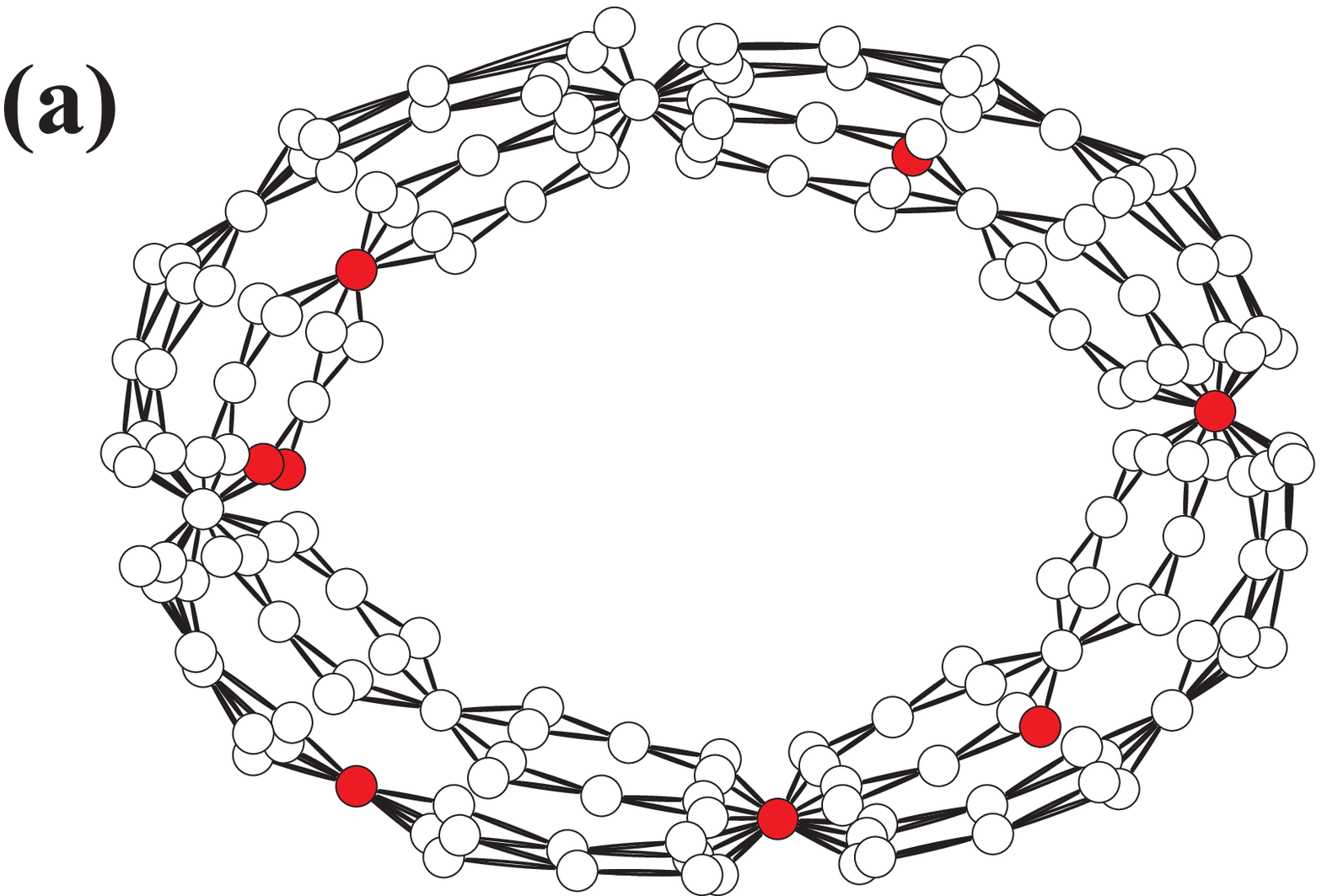}}
\scalebox{0.4}[0.4]{\includegraphics[trim=0 0 0 0]{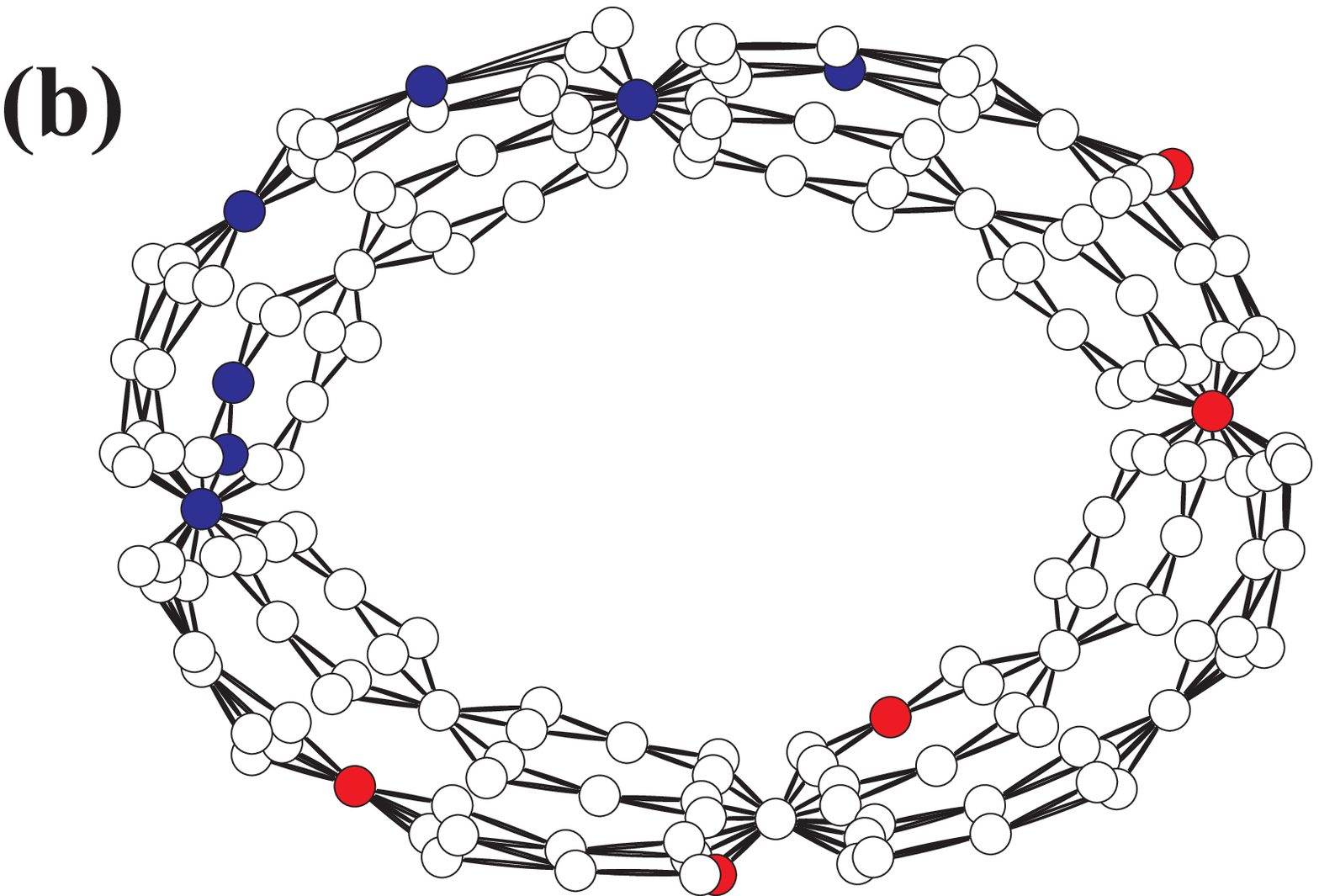}}
\caption{(Color online) Illustration of the $A+A\to\emptyset$ and
$A+B\to\emptyset$ processes on the IDS-SF networks with $n=4$. (a)
$A+A\to\emptyset$ at $t=10$, and (b) $A+B\to\emptyset$ at $t=15$.
Red and blue plots denote $A$ and $B$ particles respectively.}
  \label{depletion and segregation}
\end{center}
\end{figure}

As shown in Fig.~\ref{depletion and segregation}(a), a hub leads to
a fast decay of the local $A$ particle density in the beginning,
followed by a slow decay in the long time regime as shown in
Fig.~\ref{AA_q=0}(a). Thus, one can clearly observe depletion zones
emerging from the intervals among hubs in this panel. In
Fig.~\ref{AA_q=0}(b), a hub in a $A$ or $B$-rich domain can give
rise to a pure $A$ or $B$ zone after a prompt local annihilation of
$A$ and $B$, leaving a relatively particle-free space among the
hubs. With these segregations, one can observe a slow decay of the
reaction rate as shown Fig.~\ref{AB_q=0}(a).

\begin{figure}
\begin{center}
\scalebox{1}[0.9]{\includegraphics[trim=0 0 0 0]{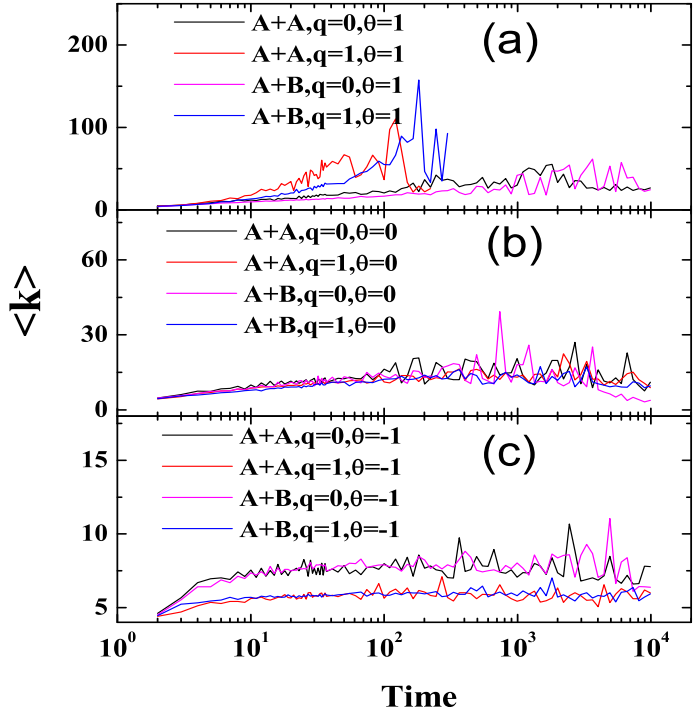}}
\caption{(Color online) $\langle k\rangle$ as a function of time $t$
for $A+A\to\emptyset$ and $A+B\to\emptyset$ for $\theta=-1,0,1$ with
$q=0,1$ respectively.}\label{AVE_K}
\end{center}
\end{figure}

Interestingly, the depletion zone and segregation also inhibit
particles moving from leaves to hubs when $\theta<0$. The behavior
can be observed by measuring the average degree of occupied nodes
increases with $t$ as plotted in Fig.~\ref{AVE_K}. Note that the
plots are also logarithmically binned. Recalling the discussion at
the beginning of this section, particles are attracted by the leaves
in this condition. For $q=1$, particles tend to agglomerate around
hubs for $\theta>0$ (see Fig.~\ref{AVE_K}(a)) and leaves for
$\theta<0$ (see Fig.~\ref{AVE_K}(c)). Comparing Fig.~\ref{AA_q=1}(a)
and Fig.~\ref{AB_q=1}(a) with Fig.~\ref{AA_q=0}(a) and
Fig.~\ref{AB_q=0}(a), one can figure out the depletion zone and
segregation slow down the rate of reaction for $\theta>0$. For
$\theta<0$, comparing Fig.~\ref{AB_q=1}(c) with
Fig.~\ref{AB_q=0}(c), instead, they accelerate the rate slightly.
This is because they slow down the decentralizing process of the
particles. For $\theta=0$, their effect is hardly identified as
shown in Fig.~\ref{AVE_K}(b) without the enhancement of weight.
Apparently, these interesting behaviors are vastly distinct from the
previously reported results~\cite{NJP11063025,PRE82021108}.

\subsection{Case of $0<q<1$}
For $0<q<1$, the networks are stochastic, which makes them not
self-similar~\cite{PRE80061111}. Thus the networks are non-fractal
in this middle case. Thus, in order to discuss the variation in the
dependence of $f$ on $q$, we have performed extensive numerical
simulations for various $q$ from $0$ to $1$. The simulation settings
were the same as the former cases. When $q$ increasing from $0$ to
$1$, the exponent of global mean first-passage time of random walks
$G(N_g)$, decreases from $1$ to
$\frac{\ln3}{\ln4}$~\cite{PRE80061111}, which indicates the
enhancement of transporting efficiency during the process. At the
same time, the diameter of the networks also decreases while the
disassortative mixing feature disappears.

\begin{figure}
\begin{center}
\scalebox{1}[0.9]{\includegraphics[trim=0 0 0
0]{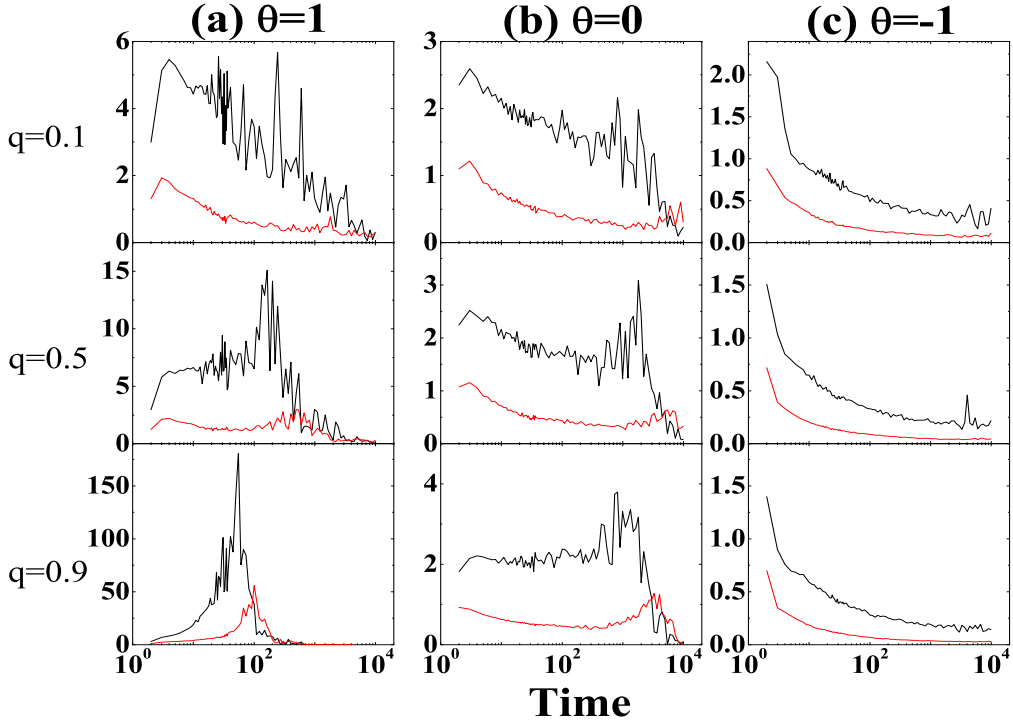}} \caption{(Color online) Reaction rate
exponent $f$ with $\theta=-1,0,1$ and $q=0.1,0.5,0.9$ for
$A+A\to\emptyset$ (black lines) and $A+B\to\emptyset$ (red lines)
processes on the IDS-SF networks respectively.}
  \label{A+A_A+B_0.1-0.9}
\end{center}
\end{figure}

In Fig.~\ref{A+A_A+B_0.1-0.9}, one can observe that the segregations
among hubs disappear gradually with the increase of $q$. Under the
influence, the diffusion rate increases drastically, leading to an
apparent enhancement of $f$ for $\theta=1$ (see
Fig.~\ref{A+A_A+B_0.1-0.9}(a)). Notably, in panel (b), the purely
topological segregations for $\theta=0$ are also observable,
although its influence is not as apparent as that in panel (a).
Also, comparing $q=0.1$ with $q=0.9$, the subtle influence of
segregations on the reaction rate in the case of $\theta=-1$ can be
identified in panel (c). As shown in this panel, the reaction rate
decreases slightly with $q$, which is consistent with our
observation in Section~\ref{CASE 0} .

\smallskip

\section{CONCLUSION}\label{sec:conclusion}
In summary, we have investigated the diffusion-annihilation process
on a family of weighted scale-free networks with identical degree
sequence (weighted IDS-SF networks), which is controlled by a
parameter $q\in[0,1]$. In this paper, the weight of links is defined
as $w_{ij}=(k_ik_j)^{\theta}$ with the degree $k_i$ and $k_j$ of
both nodes, where $\theta$ is the network's weightiness parameter.
For a convenience, we define a kinetic exponent $f$ as
$\frac{d(\frac{1}{\rho(t)})}{dt}$, where $\rho(t)=\rho_A(t)$ for the
$A+A\rightarrow \emptyset$ process and $\rho(t)=\rho_A(t)+\rho_B(t)$
for the $A+B \rightarrow \emptyset$ process. Based on the
definition, we provide numerical results to characterize the
relation between $f$ and the reaction time $t$ for the
$A+A\rightarrow \emptyset$ and $A+B \rightarrow \emptyset$
bimolecular reactions.

One significant observation is that, in contrast to the commonly
accepted conception that the depletion zone and segregation only
exist in fractal networks. Our observation shows they can exist in
the diffusion-annihilation process on the non-fractal networks as
well. This striking feature in scale-free networks was not reported
in previous studies. In fact, the depletion zone and segregation can
both exists in fractal and non-fractal networks no matter whether it
is weighted or not. We found that the segregation effect is
essentially caused by the disassortative mixing, i.e., high-degree
nodes tend to connect with low-degree nodes. On the weighted
networks, its influence on the particles diffusion is highly
enhanced by the weight heterogeneity. We have demonstrated that both
degree and weight distribution do not suffice to characterize the
diffusion-annihilation processes on weighted scale-free networks.
Our observations suggest care should be taken when making general
statements about the diffusion-annihilation process in weighted
scale-free networks.

\smallskip

\section*{Acknowledgment}
This work was supported by National Natural Science Foundation of
China (NSFC) under grants Nos. 60873040, 60873070, 61074119, and the
Shuguang Program of Shanghai Municipal Education Commission and
Shanghai Education Development  Foundation.

\end{document}